# Preprint notes

## Title of the article:

Implementing Ethics in AI:

Initial Results of an Industrial Multiple Case Study

## Authors:

Ville Vakkuri, Kai-Kristian Kemell, Pekka Abrahamsson

## Notes:

# Implementing Ethics in AI:

# Initial Results of an Industrial Multiple Sase study

**Abstract.** Artificial intelligence (AI) is becoming increasingly widespread in system development endeavors. As AI systems affect various stakeholders due to their unique nature, the growing influence of these systems calls for ethical considerations. Academic discussion and practical examples of autonomous system failures have highlighted the need for implementing ethics in software development. However, research on methods and tools for implementing ethics into AI system design and development in practice is still lacking. This paper begins to address this focal problem by providing elements needed for producing a baseline for ethics in AI based software development. We do so by means of an industrial multiple case study on AI systems development in the healthcare sector. Using a research model based on extant, conceptual AI ethics literature, we explore the current state of practice out on the field in the absence of formal methods and tools for ethically aligned design.

# 1 Introduction

The role of ethics in software systems development has dramatically changed following the increasing influence of Autonomous Systems (AS) and Artificial Intelligence (AI) systems. AI/AS systems necessitate ethical consideration due to their unique nature. Whereas one can opt out of using conventional software systems, the very idea of being an active user in the context of AI systems is blurred.

The harm potential of these systems, as well as actual real-life incidents of AI system failures and misuse, have resulted in a growing demand for AI ethics as a part of software engineering (SE) endeavors. AI ethics studies have argued that AI/AS engineering should not be simply a technological or an engineering endeavor [1]. Specifically, it is argued that developers should be aware of ethics in this context due to their key role in the creation of the systems. Aside from discussion among the academia, public voices have also expressed concern towards unethical AI systems following various real-life incidents (e.g. unfair systems [2]).

Despite the increasing activity on various fronts in relation to AI ethics, a notable gap between the concerns voiced over AI ethics and SE practice in AI remains. It is known that developers are not well-informed on ethics [3]. New ethical methods and practices that take into account the behavioral and social aspects of SE are needed. Thus, AI Ethics also needs to be approached from the field of behavioral software engineering (e.g. [4]). Developers are known to prefer simple and practical methods, if they utilize any at all [5], which makes the lack of methods in AI ethics an issue. Without methods, it can be difficult for organizations to detect ethical issues during design and development, which can become costly later on.

Extant studies on AI Ethics have largely been theoretical. To provide empirical data into this on-going discussion on AI ethics, we have conducted a multiple case study on AI system development in the healthcare sector in order to further our understanding on the current state of practice. The exact research question tackled here is:

**RQ:** How are AI ethics taken into consideration in software engineering projects when they are not formally considered?

# 2 Related work

Much of the research on AI ethics has been conceptual and theoretical in nature. These studies have e.g. focused on defining AI ethics in a practical manner through various constructs in the form of values. For the time being, this discussion on defining AI ethics has come to center around four values: transparency [6-8], accountability [6,8], responsibility [6,8], and fairness (e.g. [2]), as we discuss in the next section.

Following various real-life incidents out on the field, AI ethics has also begun to incite public discussion. This has caused various government, public, and private organizations to react, primarily by producing their own demands and guidelines for involving ethics into AI development. Countries such as Germany [9] have emphasized the role of ethics in AI /AS, and the EU drafted its own AI ethics guidelines [10]. Industry organizations such as Google and IBM[1] have also devised their own guidelines.

Thus far, various attempts to bring this on-going academic discussion out on the field have been primarily made in the form of guidelines and principles, with the most notable ones being the IEEE guidelines for Ethically Aligned Design (EAD) [8]. However, past experiences have shown us that guidelines and principles in the field of ICT ethics do not seem to be effective. For example, McNarama [3] argued based on empirical data that the ACM ethical guidelines had ultimately had very little impact on developers, who had not changed their ways of working at all. A recent version of the EAD guidelines acknowledged that this is likely to also be the case in AI ethics.

[1] Google: AI Principles: https://www.blog.google/technology/ai/ai-principles/
IBM: Everyday ethics for AI: https://www.ibm.com/watson/assets/duo/pdf/everydayethics.pdf

## 3 Research Model

The field of AI ethics can be divided into three categories: (1) Ethics by Design (integration of ethical reasoning capabilities as a part of system behavior e.g. ethical robots); (2) Ethics in Design (the regulatory and engineering methods); and (3) Ethics for Design: (codes of conduct, standards etc.) [11]. In this paper, we focus on the ethically aligned development process.

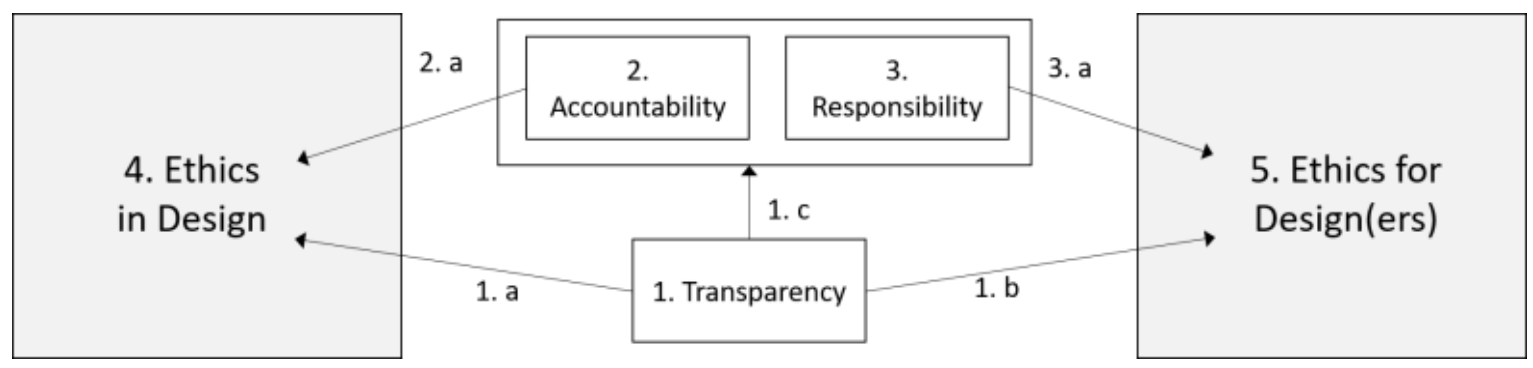

**Fig. 1.** Research framework

In addressing ethics as a part of the development of AI and AI-based systems, various principles have been discussed in academic literature. For the time being, the discussion has centered on four constructs: Transparency [6-8] Accountability [6, 8], Responsibility [6], and Fairness (e.g. [2]). A recent EU report [10] also discussed Trustworthiness as a value all systems should aim for, according to its authors. Out of these four main principles, we consider accountability, responsibility, and transparency (ART principles), as formulated by Dignum [6], a starting point for understanding the involvement of ethics in ICT projects.

Transparency is defined in the ART principles of Dignum [6] as transparency of the AI systems, algorithms and data used, their provenance and their dynamics. I.e. the transparency refers to understanding how AI systems work by being able to inspect them. Transparency can be argued currently to be the most important of these principles or values in AI ethics. Turilli and Floridi [7] argue that transparency is the key pro-ethical circumstance that makes it possible to implement AI ethics at all. It has also been included into the EAD guidelines as one of the key ethical principles [8].

In the research framework of this study, transparency is considered on two levels: (a) transparency of data and algorithms (line 1.a), as well as, (b) systems development (line 1.b). The former refers to understanding the inner workings of the system in a given situation, while the latter refers to understanding what decisions were made by whom during development. As a pro-ethical circumstance, transparency makes it possible to assess accountability and responsibility (line 1.c).

Accountability refers to determining who is accountable or liable for the decisions made by the AI. Dignum [6] in their recent works defines accountability to be the explanation and justification of one's decisions and actions to the relevant stakeholders. In the context of this research framework, accountability is used not only in the context of systems, but also in a more general sense.

Transparency is required for accountability (line 1.c), as we must understand why the system acts in a certain fashion, as well as who made what decisions during development in order to establish accountability. Whereas accountability can be considered to be externally motivated, closely related but separate construct responsibility is internally motivated. The concept of accountability holds a key role in aiming to prevent misuse of AI and in supporting wellbeing through AI [8].

Dignum [6] defines responsibility in their ART principles as a chain of responsibility that links the actions of the systems to all the decisions made by the stakeholders. We consider it to be the least accurately defined part of the ART model, and thus have taken a more comprehensive approach to it in our research framework. According to the EAD guidelines, responsibility can be considered to be an attitude or a moral obligation for acting responsibly [8]. A simplified way of approaching responsibility would be for a developer to ask oneself e.g. “would I be fine with using my own system?”. While accountability relates to the connection between one’s decisions and the stakeholders of the system, responsibility is primarily internal.

# 4 Study Design

This study was carried out as a multiple case study featuring three different cases where AI systems were developed for the needs of the healthcare sector. Each case was a specific AI project in a case company. All three projects were development projects focused on creating a prototype of an AI-based healthcare software solution. The combination of AI solutions used were different in each case. NLP (natural language processing) technologies played major role in cases B and C.

**Table 1.** Case Information

| Case | Case Description | Respondents [Reference] |
|---|---|---|
| A | Statistical tool for detecting social marginalization | Data Analyst [R1], Consultant [R2], Project Coordinator [R3] |
| B | Speech recognition and NLP based tool for diagnostics | Developer [R4], Developer [R5], Project Manager [R6] |
| C | NLP based tool for indoor navigation | Developer [R7], Developer [R8] |

The interviews were conducted as semi-structured, qualitative interviews, using a strategy prepared according to the guidelines of Galletta [12]. The interviews were conducted face-to-face and the audio was recorded. The records were then transcribed for the purpose of data analysis. In the transcripts, the cases and respondents were given individual references shown in Table 1. The interviews were conducted in Finland, using the Finnish language. The interview questions in their entirety can be found in an external resource[2]. We focused on the developer and project point of view by primarily interviewing developers and project managers.

The data from the transcripts were analyzed in two phases. First, we followed a grounded theory (Strauss and Corbin [13] and later Heath [14]) inspired approach. In this phase, the transcripts were coded quote by quote and each quote was given a code describing its contents. The same process was repeated for all eight interviews. In the second phase, we utilized the commitment net model of Abrahamsson [15], as analysis tool to further analyze and categorize the coded quotes from the first phase.

In using the commitment net model, we followed a similar method as in Vakkuri et al. 2019 [16] and focused on the concerns and actions of the developers in relation to software development. Each concern and any actions related to it were listed for each respondent and compared across respondents and cases.

These findings were then compared to the constructs in our research framework in order to evaluate what aspects of AI ethics were being implemented in the project. In this evaluation, actions were emphasized due to the research question of this study. I.e. we wanted to understand how they had implemented ethics in practice. In presenting our results, we present our key findings as primary empirical conclusions, PECs.

[2] http://users.jyu.fi/~vimavakk/AIDevQuestionnaire

# 5 Empirical Results

As the interviews progressed, the developers expressed some concerns towards various ethical issues. However, these concerns were detached from their current work. Furthermore, it was evident that in none of the cases had the hypothetical effects of the system on the stakeholders been discussed. To give a practical example, a system potentially affecting memory illness diagnoses (Case B) clearly has various effects on its potential users when the test can be taken without supervision. Yet, the developers of this system felt that their users would not be curious about the workings of the system. They considered it sufficient if the responsibility was outsourced to the user and it was underlined that the system does not make the formal diagnosis.

The developers also exhibited a narrow view of responsibility in relation to harm potential. Only physical harm potential was considered relevant, and the developers felt that none of their systems had such potential.

"Nobody wants to listen to ethics-related technical stuff. […] It's not relevant to the users" (R5)

"What could it affect… the distribution of funds in a region, or it could result in a school taking useless action… it does have its own risks, but no one is going to die because of it" (R1)

**PEC1** Responsibility of developers and development is under-discussed.

In terms of transparency of algorithms and data, case A stood out with the team's mathematical knowledge. They utilized algorithms they were familiar with and which they understood on an in-depth level. In cases B and C, the companies utilized third-party components largely as black boxes. They did, however, have an in-depth understanding of any components created by the team. Even though transparency of algorithms and data was not present, case B developers acknowledged its potential importance. However, it was not pursued in projects B and C due to not being a formal requirement, as opposed to A where it was pursued due to being one.

"We have talked about the risks of decision-making support systems, but it doesn't really affect what we do" (R5)

**PEC2** Developers recognize transparency as a goal, but it is not formally pursued

Accountability was actively considered in relation to cybersecurity and data management, as well as error handling related to program code. The developers were aware that they were in possession of sensitive data, and that they were accountable for taking measures to keep it secure, as well as to abide to laws related to personal data handling. To this end, cybersecurity was considered as a part of standard company protocol, following established company practices.

"It's really important how you handle any kind of data… that you preserve it correctly, among researchers, and don't hand it out to any government actors. […] I personally can't see any way to harm anyone with the data we have though." (R2)

Developers' concerns on error handling, underlined by one of the respondents directly remarking "I aim to make error free software" (R1), also stood out. The developers were concerned about engineering quality software in terms of it being error free and considered it their professional responsibility to do so. The respondents could discuss various practices they utilized to handle and prevent errors in the project.

**PEC3** Developers feel accountable for error handling on programming level and have the means to deal with it

Through ethics were not taken into consideration on a project level, the individual developers exhibited some concern towards socioethical issues arising from their systems. While they were able to think of ways their system could negatively affect its users or other stakeholders in its current state, they lacked ways to address these concerns, as well as ways to conduct ethical analysis. Some extant SE practices such as documentation and audits were discussed as ways to produce transparency, but ultimately they offered little help in systematically implementing ethics.

**PEC4** While the developers speculate potential socioethical impacts of the resulting system, they do not have means to address them.

# 6 Discussion

On a general level, our findings further underline a gap between research and practice in the area. Whereas research on AI ethics alongside various guidelines devised by researchers and practitioners alike has discussed various ethical goals for AI systems, these goals have not been widely adopted out on the field.

Extant literature has highlighted the importance of transparency of algorithms and data [6, 8]. Without understanding how the system works, it is e.g. impossible to establish why it malfunctioned in a certain situation in order to understand the causes of an accident that resulted in material damages. Our findings point towards transparency being largely ignored as a goal. Third-party components are utilized as black boxes, and developers do not see this as a notable problem. In this sense, we consider PEC2 to contradict existing literature. The lack of emphasis placed on transparency is interesting from the point of view of feature traceability as well. For decades, understanding the inner workings of a system was considered important in any SE endeavor [17]. In AI SE, this long-standing goal of feature traceability seems to be waning.

The situation is similar for tackling potential misuse of the systems, error handling during system operations, and handling unexpected system behavior. These goals are included into the IEEE EAD guidelines [8]. Yet, none of the case companies took any measures to address these potential issues. Error handling was simply considered on the level of program code. To this end, though we discovered various examples of ethics not being implemented, we also discovered that some existing and established SE practices can be used to support the implementation of AI ethics. Documentation, version control, and project management practices such as meeting transcripts produce transparency of systems development by tracking actions and decision-making. Similarly, software quality practices help in error handling in the context of AI ethics (PEC3), although only on the level of program code.

The developers exhibited some ethical concerns towards the systems they were developing (e.g. PEC2). Little is currently known about the state of practice out on the field, although a recent version of the EAD guidelines speculated about a gap in the area, which our findings support in relation to most aspects of AI ethics. Despite AI ethics largely not being implemented, our findings point towards it partially being a result of a lack of formal methods and tools to implement it (PEC4).

Thus, following this study, as well as a past case study [16], we suggest that future research seek to tackle the lack of methods and tooling in the area. Though developers may be concerned about ethical issues, they lack the means to address these concerns. Methods can also raise awareness of ethics, motivating new concerns.

As for the limitations of the study, the outlined research model is heavily based on ART principles of Dignum [6] and IEEE's EAD [8]. This may exclude some parts of the current AI ethics discussion (e.g. Fairness). However, the EAD can be seen as a distilled version of the ongoing AI ethics discussion that includes the most important parts of it. Finally, the sample size is quite small for making far reaching conclusions but provides much needed empirical data on a very current topic.

# 7 Conclusions & Future work

In this paper, we have sought to better understand the current state of practice in AI ethics. Specifically, we studied the way AI ethics are implemented, if at all, when they are not formally considered in a software engineering project. To this end, we conducted a multiple case study featuring three case companies developing AI solutions for the healthcare sector.

We discovered that some existing good practices exist for some aspects of AI ethics. For example, current practices out on the field are already capable of producing transparency of systems development. Moreover, the developers are aware of the potential importance of ethics and exhibit some concerns towards ethical issues. Yet, they lack the tools to address these concerns. As tackling ethics is not a formal requirement in AI projects, these concerns go unaddressed for business reasons. In this light, we consider the creation of methods and tools for implementing AI ethics important. These will both help developers to implement AI ethics in practice as well as raise their awareness of ethical issues by e.g. helping them understand harm potential of AI systems.